\documentclass[letterpaper,twocolumn,10pt]{article}
\usepackage{usenix,epsfig,endnotes}

\usepackage[utf8]{inputenc}
\usepackage{times}
\usepackage{graphicx}
\usepackage{macros}
\usepackage{hyperref}

\usepackage{subcaption}
\usepackage{listings}
\usepackage{xcolor}
\usepackage{xspace}
\usepackage{lastpage}
\usepackage{booktabs}
\usepackage{paralist}
\usepackage[noend]{algpseudocode}
\usepackage{algorithm}
\let\oldReturn\Return
\renewcommand{\Return}{\State\oldReturn}

\usepackage[export]{adjustbox}

\newcommand{\name}{IPA}

\newtheorem{definition}{Definition}
\newtheorem{theorem}{Theorem}

\lstdefinestyle{basic}{
    basicstyle={\scriptsize\linespread{0.6}\ttfamily},
    numbers=left,
    numberstyle=\tiny\color{gray}\ttfamily,
    numbersep=5pt,
    backgroundcolor=\color{white},
    showspaces=false,
    showstringspaces=false,
    showtabs=false,
    frame=noe,
    rulecolor=\color{black},
    captionpos=b,
    keywordstyle=\color[rgb]{.4,.4,.6}\bf,
    commentstyle=\color{gray},
    stringstyle=\color[rgb]{.3,.5,.3}\bf,
    keywordstyle={[2]\color{black}\bf},
    tabsize=1,
}

\lstdefinelanguage{customJava}
{
    morekeywords={@forall,@Invariant,@Inv,@True,@False,@Increments,@Decrements,@Counter,forall,and},
    sensitive=false,
    morecomment=[l]{//},
    morestring=[b]",
    morestring=[b]',
}

\newcommand{\includecode}[2][c]{\lstinputlisting[escapechar=, style=basic,language=Java]{#2}}

\usepackage{titlesec}

\titlespacing\section{0pt}{6pt plus 2pt minus 2pt}{1pt plus 1pt minus 1pt}
\titlespacing\subsection{0pt}{4pt plus 2pt minus 2pt}{1pt plus 1pt minus 1pt}
\titlespacing\subsubsection{0pt}{3pt plus 1pt minus 1pt}{1pt plus 1pt minus 1pt}

\begin{document}

\date{}

\title{IPA: Invariant-preserving Applications for Weakly-consistent Replicated Databases}

\author{
{\rm Valter Balegas, Nuno Preguiça}\\
{\rm Sérgio Duarte, Carla Ferreira}\\
NOVA LINCS \& DI, FCT,\\ Universidade NOVA de Lisboa
\and
{\rm Rodrigo Rodrigues}\\
INESC-ID \& IST, University of Lisbon
}

\maketitle

\thispagestyle{empty}

\begin{abstract}


Storage systems based on \WC{} provide better availability and
lower latency than systems that use \SC{}, especially in
geo-replicated settings.
However, under \WC{}, it is harder to ensure the
correctness of applications, as the execution of uncoordinated operations
may lead to invalid states.

In this paper we show how to modify an application to make it
run correctly under \WC{}.
We developed an analysis that detects which operations
need to be corrected, and proposes possible modifications
to operations to prevent inconsistencies.
This analysis
allows the programmer to choose the preferred semantics for each
problematic execution, while preserving the original semantics of
operations when no conflicts occur.
The modified application runs with small overhead when compared
with its \WC{} counterpart, which cannot preserve application correctness.

%

\end{abstract}

\section{Introduction}

To meet the expectations of their users, global scale services
need to be highly available, scalable, and fast. Failing to address
these requirements may harm revenue and ultimately lead to the shutdown of the service~\cite{Schurman09latency}.

The key technique to achieve these properties is replication~\cite{pnuts,spanner,dynamo,cops,eiger,redblue,walter}.
Some replicated systems adopt a \SC{} model, enforcing a single view of
the data across different replicas, which simplifies application development.
While several designs exist for scaling \SC{}
systems~\cite{farm,ramcloud,spanner}, they only provide low
latency within the boundaries of a data center,
since replicas need to coordinate before replying to each request.
As a consequence, many systems forego \SC{} to
achieve low latency and high availability in the presence of partitions.

\WCc{} models~\cite{cops,eiger,swiftcloud,dynamo,cassandra} allow
replicas to diverge temporarily by accepting updates in a given replica and
executing them locally without coordinating with other replicas.
After the execution finishes, updates are propagated asynchronously to other replicas.
This provides low latency and high availability, but also allows the replicated system to expose anomalous states that are not allowed by \SC{}.
This can happen whenever the effects of an operation submitted at some replica
are applied at some other replica in a state where the operation preconditions
no longer hold, even when convergence policies ensure state-convergence~\cite{bailisiser}.
In this case, the final state can violate an application invariant.
Identifying and addressing these potential problems is complex, with services often
going into production with errors that might break the
intended application semantics~\cite{feralcc}.

A middle ground approach that tries to approximate the best of
both worlds consists in splitting operations into a subset
that can execute under \WC{} and another
subset that requires coordination~\cite{redblue,indigo,cise,Roy14Writes,bailisiser}.
In this case, operations that can potentially lead to an invariant violation will incur
in a coordination cost several orders of magnitude
higher than the cost of running that same operation
in a single data center.


In this paper, we explore a different route for achieving the holy grail of fast and invariant-preserving operations. In our new approach, instead of coordinating the pairs of invariant-violating operations at runtime, we modify the application logic at development time to prevent invariant violations despite the concurrent execution of any operation. This is achieved by adding the necessary application logic for ensuring that the integration of the effects
of an operation at remote replicas always conforms to the application invariants,
despite any concurrent operations.

Since writing such logic is a complex and application-dependent task, we propose \name{}, a process to modify an application
in a way that meets this property. \name{} performs a
static analysis on the application specification to:
\begin{inparaenum}[(i)]
\item identify the operations whose preconditions can be invalidated
by the concurrent execution of other operations;
\item propose changes to those operations
to guarantee that the preconditions are restored despite any concurrent changes.
\end{inparaenum}

The first step in this process shares some goals of prior work~\cite{indigo,cise}, and therefore our design reuses their algorithms to identify
the pairs of operations that can lead to an invariant violation (in this prior work the execution of those
operations was then coordinated at runtime).
However, we significantly extend that design in two main ways. First,
\name{} needs not only to
identify such pairs, but also
the root cause of the invariant violation. Second, once that
root cause is determined, \name{} needs
to propose alternative operations that do not suffer the same problems.
We provide a complete toolset that programmers can use to achieve these steps.
The resulting applications can execute in any replicated system with \CC{},
type-specific conflict resolution, and highly available
transactions~\cite{hat}.

Our evaluation shows that the proposed approach leads to latency and
scalability similar to \WC{} models, while preserving global
application invariants. When compared with state of the art solutions
that maintain invariants by adding coordination~\cite{indigo}, no operations
experience high latency due
to the need of coordinating with other replicas.

We studied multiple real world applications that specified invariants~\cite{feralcc}
and found that, in most cases, it was possible to use \name{}
to make applications invariant preserving, while also maintaining
semantics that were acceptable from the standpoint of the application logic.

In summary, our contributions are the following.

\begin{asparaenum}[(i)]

\item A novel approach to achieve fast and invariant-preserving applications,
  based on transforming the applications to become correct when running
  under \WC{} by design;

    \item An algorithm for proposing modifications to
    applications to make them invariant-preserving;

    \item A set of novel convergent data types with the required conflict
    resolution semantics;

    \item A systematic evaluation of the proposed approach, covering both performance
    and a feasiblity study based on real world applications.

\end{asparaenum}

\section{Invariant preserving applications}
\label{sec:overview}

This section introduces the system model and defines a set of criteria
for invariant preservation.
As a running example, we will use an application to manage information about gaming tournaments.
In this application, players scattered across the globe compete with each other by enrolling in tournaments.
New players can register on the system and tournaments can be added or removed.

\subsection{System model}

We consider a database composed by a set of objects fully replicated in
multiple data centers.
An operation has an associated a piece of code that executes a sequence of reads and updates
enclosed in a transaction.
As the transaction executes, the effects of updates are recorded and queued for replication upon
transaction commit.
The propagation of updates between replicas can be asynchronous and
it must respect causal ordering.
Hereafter, when we use the term operation, it refers to the set of updates produced by the execution of the transaction code in the initial replica.

We denote by $o(S)$ the state after applying the updates of
operation $o$ to some state $S$.
We define a database snapshot, $S_n$, as the state of the database after
executing a sequence of operations $o_1,\ldots,o_n$
from the initial database state, $S_{init}$, i.e., $S_n =
o_n(\ldots(o_1(S_{init})))$.
The set of operations reflected in a database snapshot $S$ is denoted by $Ops(S)$,
e.g., $Ops(S_n) = \{o_1,\ldots,o_n\}$.
The state of a replica results from applying both local and remote
operations, in the order received.

We say that an operation $o_a$
happened-before operation $o_b$ executed initially in the database snapshot $S_b$,
$o_a \prec o_b$, iff $o_a \in Ops(S_b)$.
Two operations $o_a$ and $o_b$ are concurrent, $o_a \parallel o_b$, iff
$o_a \not \prec o_b \wedge o_b \not \prec o_a$ \cite{happensbefore}.

For an execution of a given set of operations $O$, the happens-before relation defines
a partial order among them, $\mathcal{O} = (O,\prec)$.
We say $\mathcal{O'} = (O,<)$ is a valid serialization of $\mathcal{O} = (O,\prec)$
if $\mathcal{O'}$ is a linear extension of $\mathcal{O}$, i.e., $<$ is a total order
compatible with $\prec$.

Operations can execute concurrently, with each replica
executing operations according to a different valid
serialization.
This raises the problem that the state of the various replicas of
the database may diverge, in case
these operations do not commute. To prevent this,
we assume the system gives the programmer the choice of various
deterministic conflict resolution rules
to achieve state convergence on a per-object basis, i.e., the result of
applying updates that were executed concurrently is deterministic
independently of the execution order.
In our prototype, we rely on conflict-free replicated data types (CRDTs)
\cite{crdts,walter} to achieve this goal.
A CRDT defines a data type (e.g., sets, maps, counters) with deterministic
rules for handling concurrent updates -- for example, instead of using
conventional set data types, a programmer can choose between using
an \emph{add-wins} set or a \emph{rem-wins} set. In this case,
when the same element is concurrently added and removed, after merging the concurrent
updates the element will either be maintained or removed from the set, respectively.

We consider that application correctness can be expressed in
terms of invariants \cite{bailisiser,indigo,cise}.
An invariant is a logical condition expressed over the database
state.
A given state $S$ preserves an invariant $I$ iff $I(S) = true$,
where $I(S)$ is a function that checks the validity of the invariant in state $S$.
A state $S_i$ is \emph{I-valid} (or simply valid) iff $I(S_{i}) = true$; otherwise the state is
\emph{I-invalid} (or simply invalid).
We require the initial state, $S_{init}$, to be valid.

We say that $\mathcal{O'} = (O,<)$ is an I-valid serialization of $\mathcal{O} = (O,\prec)$
if $\mathcal{O'}$ is a valid serialization of $\mathcal{O}$, and $I$ holds in every state that
results from executing any possible prefix of $\mathcal{O'}$.
If $I$ is the invariant that results from the conjunction of all application invariants, then we say that
an application is correct if, in any possible execution of that application,
every replica evolves through a sequence of I-valid serializations.

\subsection{Principles for \name{}}

We now study the conditions for having correct applications under weak consistency.
We follow the notions introduced by Bailis et al.~\cite{bailisiser}, adapting
them to the notation of our model.

\vspace{-4pt}

\begin{definition}
Given a commutative set of operations $O$ and the happens before relation among operations, $\prec$,
we say $O$ is \IConfluent{}~\cite{bailisiser} iff any state $S_i$,
obtained by executing a prefix of any valid serialization of $(O,\prec)$, starting from any \emph{I-valid} state, is \emph{I-valid}.
\end{definition}

\vspace{-4pt}

This means that for a set of \IConfluent{} operations,
despite executing operations in a different serialization order,
every replica will evolve only through \emph{I-valid} states.
Along with the commutativity of the operations, this guarantees the correctness of application execution both in terms of convergence and invariant-preservation.

To preserve invariants, an operation should only produce its expected side effects in states
that satisfy the operation preconditions.
For example, for adding a player to a tournament, the player and
tournament must exist.
When an operation executes in the initial replica, the code of the
operation verifies that the local database state satisfies the
operation preconditions, leading to a state that preserves application
invariants.

The challenge arises when operation side-effects propagate asynchronously
to remote replicas.
At the remote replica, concurrent operations emitted elsewhere may have
already executed, leading to a database state where the operation preconditions
do not hold anymore.
Applying the side-effects as-is may result in an invariant violation.
For example, should the effects of adding a player to a tournament be
applied in a state where the tournament has been removed, it would lead
to the violation of the invariant that specifies that a player can
only enroll in tournaments that exist.

We say an operation $o_1$ conflicts with $o_2$ if the execution of
$o_1$ makes the preconditions of $o_2$ false in some database state.
As such, to ensure that an operation executes correctly at a remote replica,
it is both necessary and sufficient to guarantee that the preconditions of the operation
are valid when it is executed there.


We now define in formal terms the
states in which an operation can be applied.

\vspace{-4pt}

\begin{definition}
Given a set of operations $O$ and the happens-before order among them $\mathcal{O} = (O,\prec)$,
we say that a state $S$ is an admissible state for $o \in O$ iff there is
a valid serialization of $\mathcal{O}$ in which $S$ is the state of applying
all operations that precede $o$ in the serialization order to the initial state.
\end{definition}

\vspace{-4pt}

With this definition in place, we can state a sufficient condition for
having \IConfluent{} operations, thus enabling the system to execute operations
in remote replicas without violating the invariants.

\vspace{-4pt}

\begin{theorem}
Given a set of commutative operations $O$ and the happens-before order among them $\mathcal{O} = (O,\prec)$
if for any operation $o$ and admissible
state $S$ of $o$, $o(S)$ is also an \emph{I-valid} state,
then $O$ is an \IConfluent{} set.
\end{theorem}

\vspace{-4pt}

Due to lack of space, we omit the proof, but it follows directly from
the definition of a valid serialization.

The key insight of our approach is that in most cases it is possible
to guarantee both commutativity and the sufficient property of
Theorem 1
by leveraging commutative data types whose
effects restore the operation preconditions. The choice of data types
is guided by the programmer who indicates the appropriate
conflict resolution rules for the modified objects.


For example, the operation to enroll a player in a tournament can always execute
safely if it restores the player and tournament to exist. Restoring a player can be achieved by adding (again)
the player
and using an \emph{add-wins} conflict resolution policy for the object that holds the players.
With this policy, as an add wins over a concurrent remove, the add
in the enroll will mask the effects of any concurrent remove of the same player. Likewise, adding
the tournament will protect the enroll against a concurrent removal of the tournament.




\section{\name{} recipe}~\label{sec:ipa-recipe}

For specifying an invariant-preserving version of a
given application, the programmer must execute the
following steps:

\textbf{Step 1: Specifying applications:}
The first step consists in building a specification of the application by identifying
application invariants and operation effects.
Inferring this information automatically is outside the scope of this
work~\cite{Roy14Writes,Li14Automating}.

\textbf{Step 2: Create invariant-preserving applications:}
\name{} iteratively proposes modifications to the
application, until all operations are \IConfluent{}.
First, the algorithm picks a pair of conflicting operations, if any.
Next, a list of possible modifications to make the pair safe under concurrency is presented to the programmer.
In general, each resolution strategy will have the effects of one operation prevail over the effects of the other.
The programmer is required to
choose which resolution provides the semantics that better suits the application.
If no suitable modifications exist for some conflicting pair, the unresolved conflict is flagged.
The algorithm repeats until all conflicts are resolved or flagged.

\textbf{Step 3: Modify applications:}
The output of the previous step is an updated specification of the application, stating
the conflict-resolution associated with each predicate and the effects of each
operation. The programmer can then patch the original application according to the recipe, adding the necessary effects,
which typically requires only a few additional lines of code, as detailed in Section~\ref{sec:applications:modified}.
For conflicts flagged as unsolvable by \name{}, the programmer can resort to some coordination mechanism
to avoid concurrent execution of the offending operations~\cite{redblue,indigo}.

Fully patched applications can execute in any replicated system that provides causal consistency,
highly available transactions and the necessary type-specific conflict resolution policies.
A number of systems support these features \cite{walter,swiftcloud,antidote}.


\textbf{\name{} tool:}
We implemented the \name{} tool as a proof-of-the-concept of the proposed
methodology.
Programmers interact with the tool during the analysis process to choose the
preferred resolution rules for each data-type and the preferred resolutions for
conflicting operations.



\subsection{Specifying applications}

\begin{figure}[t]
    \centering
    \hspace{0.2cm}
    \scalebox{.85}{
	    \includecode{tournament-interface-3.java}
    }
    \caption{\mbox{\tournamentApp{} application specification (excerpt).}}
    \label{fig:code-snippet}
\end{figure}

The specification conveys information about the invariant and operations' effects.
Our prototype uses first-order logic, which can express a wide variety of
properties, including all invariants typically used in relational
databses~\cite{indigo,cise,feralcc}.
We integrate the specification of applications with Java interfaces, as shown in
Figure~\ref{fig:code-snippet}.

In the specification, the invariants (lines 1 to 9) are represented by
boolean statements, and operation effects are assignments to predicates.
Such assignments can either set the value of the predicate to \TRUE{}
or \FALSE{}, or increment or decrement the value.
As an example, the effect \texttt{@True(enrolled(p,t))}, in line 21, indicates
that \enroll{} adds player $p$ to tournament $t$.
The programmer is responsible for ensuring that the implementation
respects the semantics of the predicates.


\subsection{\hspace{-1mm}\mbox{Making operations invariant-preserving}}

Algorithm~\ref{algo:main-loop} presents the logic for creating an invariant-preserving
version of an application.
We define this algorithm as a function that receives as input the invariant, $I$,
the set of operations, {\em Ops}, and a set of convergence rules, {\em CR}, defined
for each predicate by the programmer.
The algorithm only handles boolean predicates
(lines~\ref{line:startIPA} to~\ref{line:endIPA}); in
Section~\ref{sec:extensions} we explain how to extend the algorithm to support
numeric invariants.

The main loop iterates over all pairs of conflicting operations
until no more conflicts exist. For each conflicting pair
(line~\ref{line:find}), the algorithm replaces
the initial operations specification (line~\ref{line:replace}) with the new
specification
that solves the identified conflict (line~\ref{line:repairInicial}).
If there are no alternative safe operations for the conflicting pair
with a given set of convergence rules,
the pair is flagged as unsolvable and the algorithm continues, ignoring
that pair in subsequent iterations.

\begin{algorithm}[t]
    \small
    \centering
    \algnewcommand{\LineComment}[1]{\Statex \(\triangleright\) {\small #1}}
    \begin{algorithmic}[1]
    \LineComment{\name{} main loop.}
    \vspace{-0.3mm}
    \Function{\name{}}{I, Ops, CR}\label{line:startIPA}
    	\While{ existsConflictingPair(I, Ops, CR)}
        \State{opPair $\leftarrow$  findConflictingPair(I, Ops, CR)}\label{line:find}
        \State{newPair $\leftarrow$ repairConflicts(I, opPair, CR)}\label{line:repairInicial}
        \State{Ops.replace(opPair, newPair)}\label{line:replace}
    	\EndWhile
		\vspace{-1mm}
        \Return{Ops}
        \EndFunction\label{line:endIPA}

        \Statex
        \vspace{-3.5mm}
		\LineComment{Extended conflict detection algorithm.}
		\vspace{-0.3mm}
        \Function{isConflicting}{I, OpPair, CR}\label{line:detection}
    	\If{opposingEffects(OpPair)} \label{line:opposing}
			\State{newOpPair $\leftarrow$ apply(OpPair, CR)}
	        \Return{checkConflicting(I, newOpPair, CR)}\label{line:postconflict}
		\Else
			\vspace{-1mm}
	        \Return{checkConflicting(I, OpPair, CR)}
		\EndIf
        \EndFunction

        \Statex
        \vspace{-3.5mm}
 		\LineComment{\name{} algorithm for repairing conflicts.}
		\vspace{-0.3mm}
                \Function{repairConflicts}{I, OpPair, CR} \label{line:repair}
        \State{sols $\leftarrow \emptyset$}
        \State{invPreds $\leftarrow \{$getPreds(i)$ \ \mid\ $ i $\in$ invClauses(I, opPair)$\}$} \label{line:invfor}
        \State{newOpPairs $\leftarrow$ generate(invPreds, I, OpPair)}\label{line:generatecall}
        \For{opPair $\in$ newOpPairs }
           \If{not isPairSubset(opPair, sols)}\label{line:prefix}
                \If{ not isConflicting(I, opPair, CR))}\label{line:solve}
                    \State{sols $\leftarrow$ sols $\cup\  \{ $ opPair $\}$}
                \EndIf
           \EndIf
        \EndFor
        \vspace{-1mm}
        \Return{pickResolution(sols)}\label{line:pick}
        \EndFunction

        \Statex
        \vspace{-3.5mm}
		\LineComment{New operation generation.}
		\vspace{-0.3mm}
        \Function{generate}{invPreds,  I, (op1, op2)} \label{line:generate}
        \State{seed $\leftarrow \{ p(\mathit{true})
        		\cup p(\mathit{false}) \mid p \in \mbox{invPreds}\}$}\label{line:truefalse}
        \State{effectSets $\!\leftarrow$ powerSet(seed)}
        \State{pairs $\leftarrow \emptyset$}
        \For{ p1 $\in$ effectSets}\label{line:newops:start}
        \State{pairs $\leftarrow$ pairs $\cup \{ ($newOp(op1,p1)$,$op2$)\}$}\label{line:op1}
        \State{pairs $\leftarrow$ pairs $\cup \{ ($op1$,$newOp(op2,p1)$)\}$}\label{line:op2}
        \EndFor\label{line:newops:end}
        \vspace{-1mm}
        \Return{order(pairs)}\label{line:order} \Comment{{\footnotesize by increasing no. of predicates.}}
        \EndFunction

    \end{algorithmic}
    \caption{\name{} algorithm and main functions.}
    \label{algo:main-loop}
\end{algorithm}

\begin{figure*}[t]
    \centering
    \begin{subfigure}{0.33\linewidth}
        \includegraphics[width=150px]{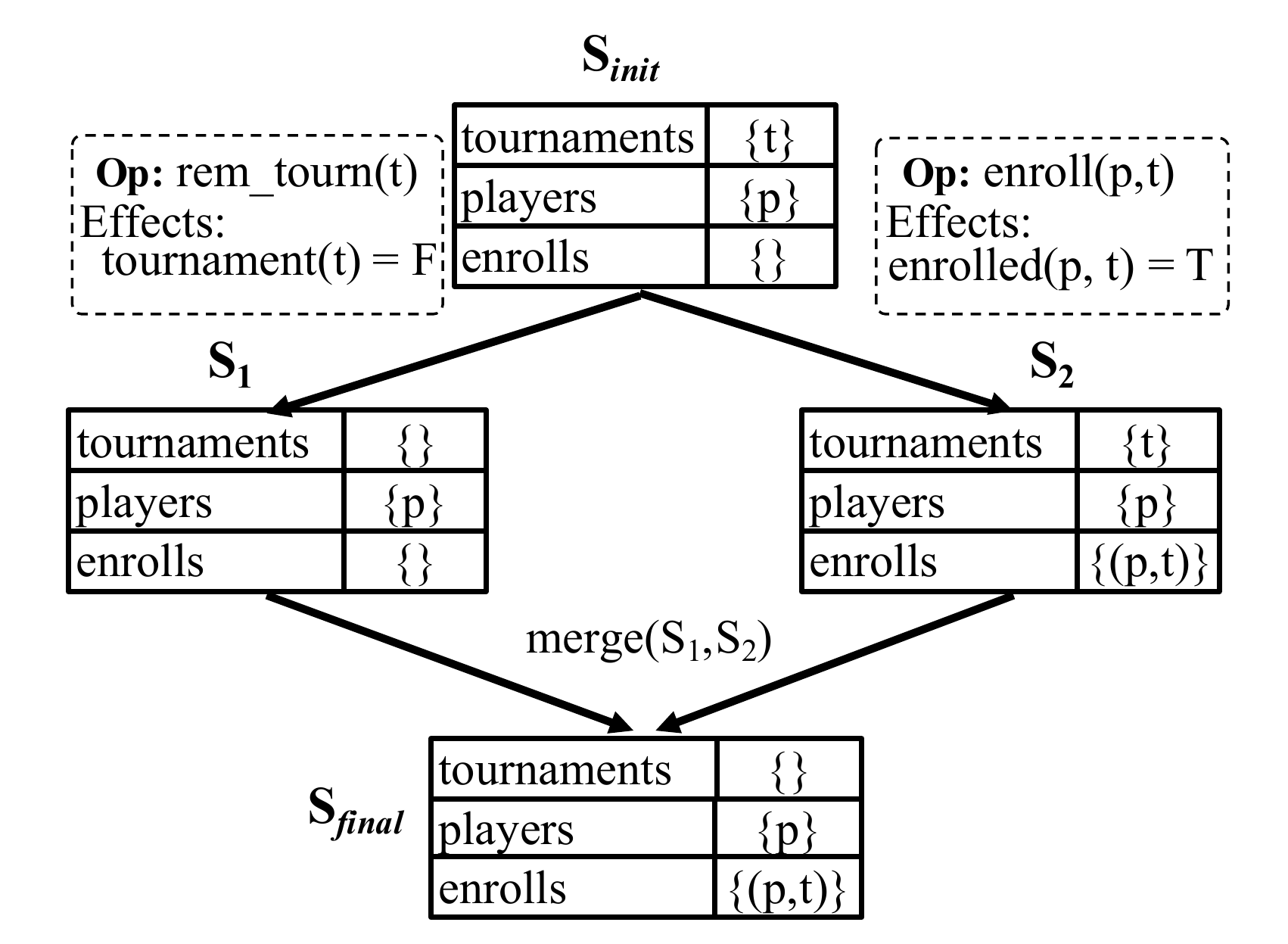}
	    \caption{Referential integrity broken.}
		\label{fig:confdetec}
	\end{subfigure}
    \begin{subfigure}{0.33\linewidth}
        \includegraphics[width=150px]{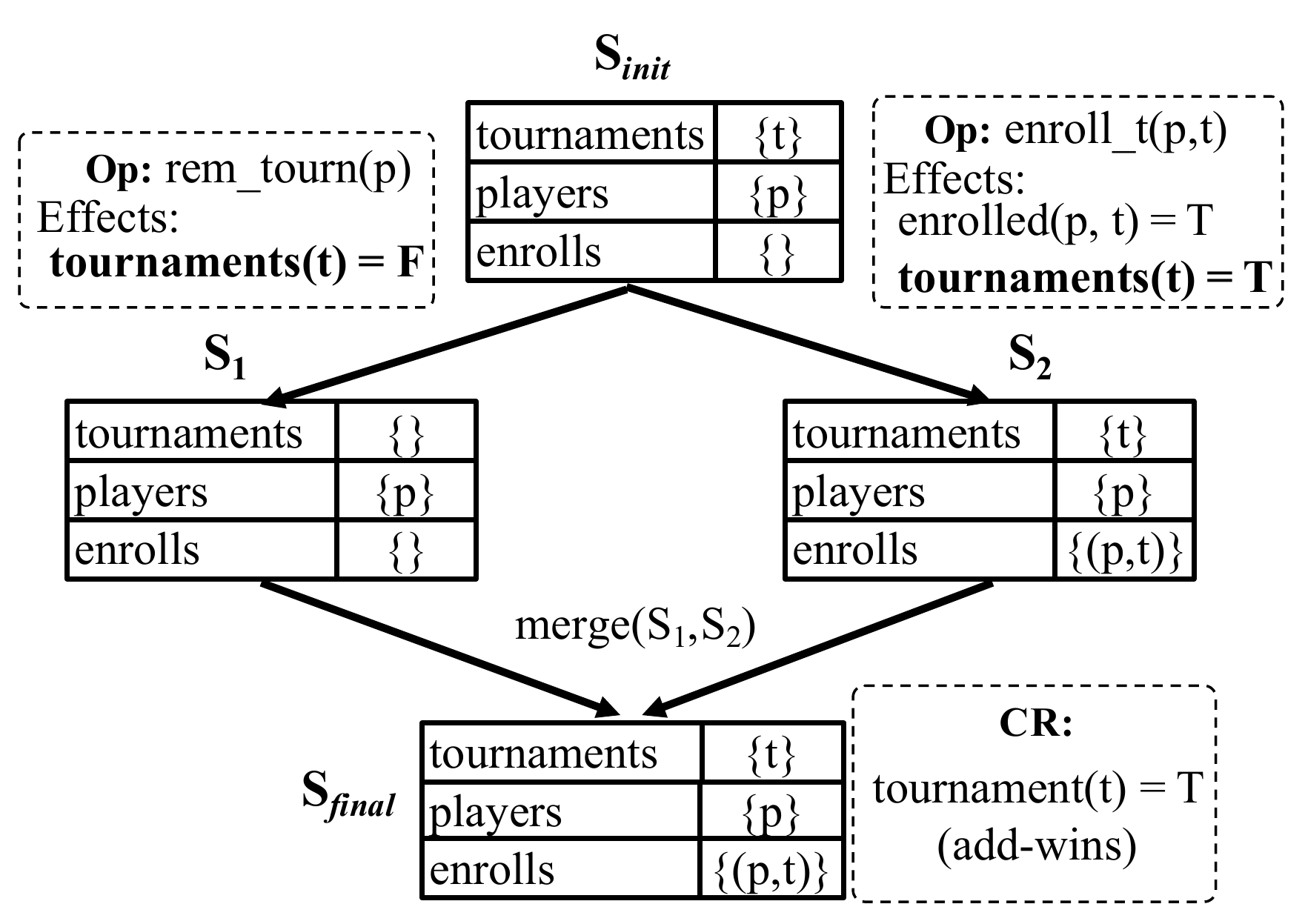}
	    \caption{Tournament is restored after merge.}
        \label{fig:enroll-p}
    \end{subfigure}
    \begin{subfigure}{0.32\linewidth}
	    \includegraphics[width=150px]{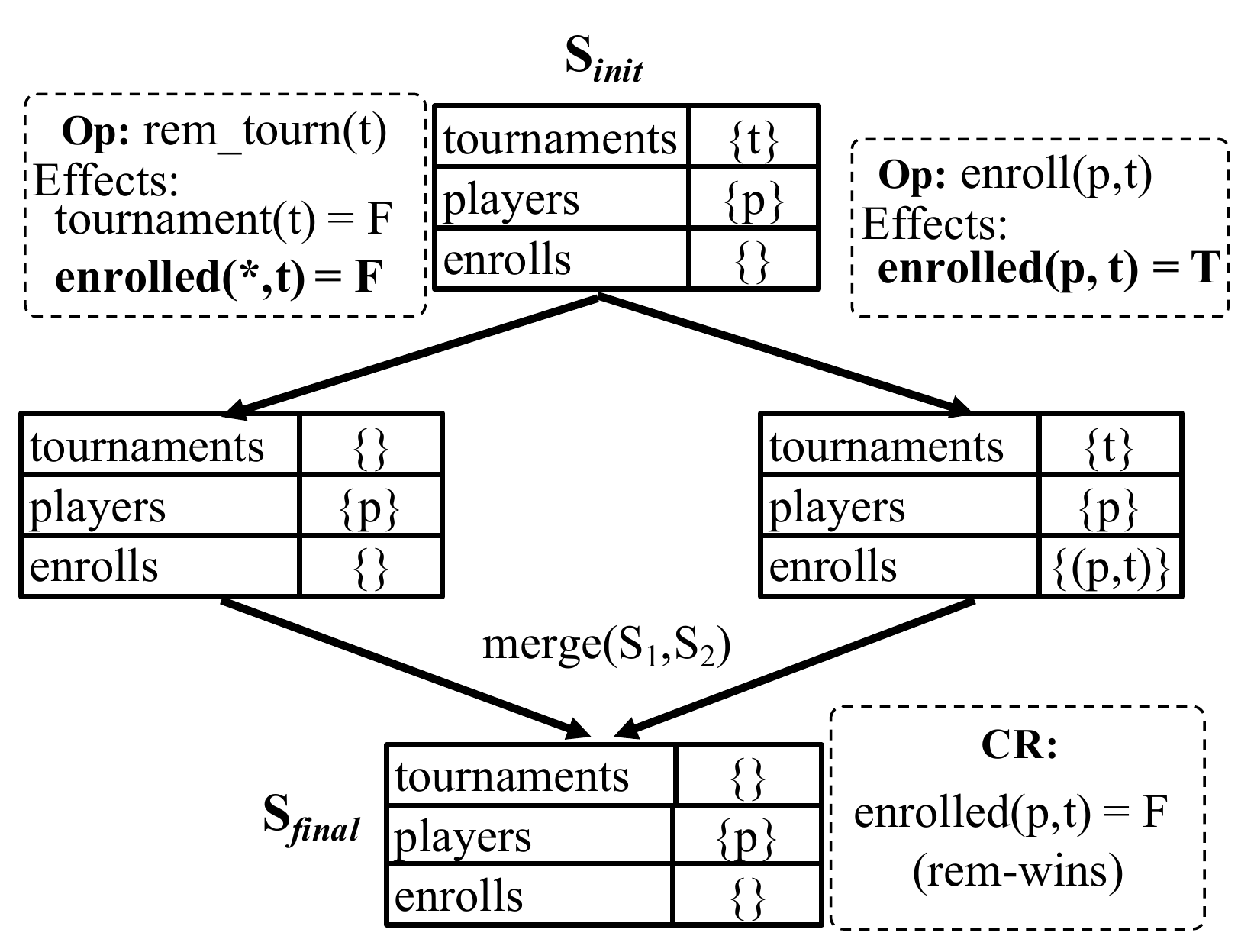}
	    \caption{Ensures player is not enrolled in any $t$.}
        \label{fig:rem-star}
    \end{subfigure}
         \vspace{2pt}
    \caption{Analysis of conflicts and resolutions of a pair of operations.}
     \label{fig:confdetec-all}
     \vspace{-10pt}
\end{figure*}

\subsubsection*{Conflict detection}

For checking conflicts, the algorithm considers all pairs of operations
(checking conflicts pairwise is sound, as shown previously~\cite{cise}).
Still, the number of test cases can be significant, thus the algorithm uses a SMT Solver to
generate all the test cases efficiently.

Figure~\ref{fig:confdetec} exemplifies the analysis of \remt{}
and \enroll{}.
The algorithm determines the weakest preconditions for executing both
operations: in this case, the predicates tournament $t$ and player $p$ must be \TRUE{}.
From $S_{init}$, the execution of each operation individually leads to
$S_1$ and $S_2$ respectively.
Combining the effects of both operation would lead to $S_{\mathit{final}}$, which violates
the referential integrity invariant stating that a player can only be enrolled
in a tournament that exists.
The algorithm identifies this pair as being conflicting,
as the precondition of each operation do not hold when the other is executed concurrently.

We extended the conflict detection mechanism proposed in Indigo \cite{indigo}
to support convergence rules.
A convergence rule, $r \in \mathit{CR}$, specifies the outcome of modifying a
predicate with opposing values: $r$ specifies that the
final value is \TRUE{} for an \addwins{} policy and \FALSE{} for a \remwins{} policy.
Supporting convergence rules in the analysis is essential for \name{} because they are
the basis for restoring operation preconditions, as we will see next.


Function {\em isConflicting} (line~\ref{line:detection}) presents
the conflict detection algorithm.
In line~\ref{line:opposing}, the function checks if the operations have
opposing effects on at least one predicate.
If so, the algorithm replaces the predicate value in each operation
with the values from the convergence rules in {\em CR}. Then, it checks
whether the combined effects of the operations
may break the invariant (line~\ref{line:postconflict}).

\subsubsection*{Proposing modified operations}

In the example, the violation can be repaired either by giving preference
to \remt{} or \enroll{}.
In the former case, it is necessary to guarantee that no player enrolls in $t$
concurrently with a \remt{}; in the latter case, that tournament $t$ is not
removed concurrently with a \enroll{}.
Our algorithm heuristically identifies the set of effects that need
to be added to each of the operations to achieve each of these behaviors,
guaranteeing that replicas converge to a correct state.


Function {\em repairConflicts} (line~\ref{line:repair}) starts by selecting, for a conflicting pair,
the invariant clauses that might be involved
in the conflict (namely the invariant clauses that have predicates affected by the effects of
the operations), and creates a pool of predicates for generating new operations (line~\ref{line:invfor}).
The next step of the algorithm is to heuristically generate new operations with combinations of
those predicates (line~\ref{line:generatecall}).
Line~\ref{line:prefix} checks if the new operations are not included in any
previous solution, ensuring
that the number of predicates added to the generated operations is minimal.
Next, the algorithm tests if the new operations solve the conflict that was identified (line~\ref{line:solve}).
All operations that solve a conflict are stored and one of them is chosen as the
resolution for the conflict, which can be done either manually or according to some policy (line~\ref{line:pick}).

The modified operations solve the conflict between the pair of operations, but they might
still conflict with other operations.
Successive iterations of the algorithm will then fix all remaining conflicts
(as said before, any unsolvable conflict is detected and flagged).

The {\em generate} function (line~\ref{line:generate}) computes all possible combinations of effects that can
be added to each operation.
The algorithm computes the powerset of predicates in {\em invPreds}, with different
predicate values \TRUE{} and \FALSE{}, and adds each element of the set to each operation,
ignoring any predicates that are already present in the operation.
The function only modifies one operation in each pair (lines~\ref{line:op1} and~\ref{line:op2}).
The generated operations are ordered by the number of predicates (line~\ref{line:order}) to ensure that the algorithm detects modified operations with fewer predicates first (in line~\ref{line:solve}).

\subsection{Example}

In this section, we analyze the solutions proposed by \name{} using the example
of the \remt{} and \enroll{} conflict.
The invariant violated by the concurrent execution of both operations is the
following: $I = $~\enrolled{} $\Rightarrow$ \player{} $\wedge$ \tournament{}.
The algorithm uses these predicates to generate new sets of effects for the operations.
We show how to modify each operation to preserve its effects over the effects
of its counterpart.

Figure~\ref{fig:enroll-p} shows \enroll{} extended with setting the predicate
\tournament{} to \TRUE{}.
When combined with an \addwins{} policy for managing tournaments, this has
the effect of recreating the tournament $t$ if a concurrent \remt{} is
executed, as shown in the final state of the figure.
We note that the additional effect has no impact if there is no concurrent
\remt{}, as the tournament has to exist for enroll{} to be executed.
This modification gives preference to the \enroll{} over \remt{}, with
the effects of the first operation prevailing while the effects of the latter
are undone.

An alternative resolution, depicted in Figure~\ref{fig:rem-star}, consists in
giving preference to \remt{} by guaranteeing that the final database state
includes no player enrolled in tournament $t$.
This can be achieved by setting the predicate $\mathit{enrolled(*,t)}$
to \FALSE{} and using a \remwins{} policy.
The wildcard ($*$) specifies that the predicate applies to any player -- this
is necessary since \remt{} has a single parameter and it is impossible to know
beforehand which players might be enrolled in tournament $t$.
With the additional effect, an \enroll{} will have no effect when
executed concurrently with a \remt{}.
In section~\ref{sec:data-types} we describe how to implement the
effect with a wildcard efficiently.

After selecting a resolution for the conflict, the algorithm proceeds
by checking if the new operations conflict with any other operations.
For instance, similar conflicts appear
when considering the pair \enroll{} and \remp{}.
Our algorithm composes the resolution of multiple operations together, until solving all
conflicts.
If the programmer is not satisfied with a set of solutions proposed by the algorithm,
he might provide a different conflict resolution set.

\subsection{Extensions}
\label{sec:extensions}
Some invariant violations cannot be prevented beforehand with a reasonable
semantics.
Take as an example the constraint in Figure~\ref{fig:code-snippet}, line 5, that
enforces a maximum number of players enrolled in a tournament.
The repair for this constraint would be to disenroll a player from the tournament,
whenever a player enrolls, which would render the application unusable.
In this case, we would like to only disenroll a player
if the size limit is actually exceeded.

Instead of applying extra effects on every operation execution,
the system can delay applying the extra effects to a later point in time,
and only do that, if a violation actually occurs.
This mechanism is known as compensation~\cite{sagas,planet,cap12years,quicksand,bayou}.
\name{} can also generate compensations, as an alternative
for modifying applications.

The analysis can automatically generate compensations for
certain constraints, like aggregation and numeric constraints, by
adding new effects as before (but executing them in a separate operation).
It additionally indicates
in which operations the compensation must be triggered.
To help the programmer, we provide data types that enforce
certain constraints using compensations out-of-the-box.
We explain the implementation of those data types in Section~\ref{sec:data-types}.

A requirement of  compensation actions is that they are commutative,
idempotent and monotonic.
This is necessary to ensure that if different replicas detect the same violations
independently, and apply a compensation action, the system still converges to
a consistent state.
\name{} only generates effects that respect these constraints.

\section{Implementation}
\label{sec:impl}
This section briefly describes the implementation of our prototype.

\subsection{\name{} tool and database support}
\label{sec:impl:tool}

The \name{} tool helps programmers writing invariant preserving applications.
The tool receives as input an annotated Java interface with the operations and
the invariants, as in the example of Figure~\ref{fig:code-snippet}, and a set
of convergence rules for predicates.
The tool runs the \name{} algorithms and
outputs the modified specification of operations,
auxiliary compensations, and the set of unsolved conflicting pairs.

The tool is implemented using Java 8.
For implementing the conflict detection and operation generation routines, we
used the Z3 SMT solver~\cite{z3} to compute test cases efficiently.

The database system that we used in our prototype
is SwiftCloud \footnote{\url{https://github.com/SyncFree/SwiftCloud}}~\cite{swiftcloud}.
SwiftCloud is a replicated key-value store with support for highly available transactions~\cite{hat},
causal consistency and per-object conflict resolution based on CRDTs~\cite{crdts}.
While transactions and causal consistency are required for correctness, the latter
feature is necessary for implementing our approach.

We implemented multiple applications for evaluation, derived from the specifications
generated by our analysis.
We use CRDT Sets as a logical representation of the predicates.
Information about predicates that involve multiple parameters are
scattered across multiple sets.

\subsection{CRDTs for supporting IPA}
\label{sec:data-types}

We now discuss the CRDTs used for implementing the resolutions
proposed by \name{}. Details on the implementation will be presented
elsewhere due to space constraints.


\subsubsection{Specialized convergence rules}

As discussed, we use CRDTs convergence rules as a key feature
to enforce invariant maintenance.
SwiftCloud already included most of the CRDTs needed to support \name{}
resolutions, such as \addwins{} and \remwins{} sets.
We needed to extend the existing set designs to support parameters that
include an wildcard (e.g. for implementing $\mathit{enrolled(*,t) = false}$).
To this end, we allow an operation (add or remove) to include a logical predicate
that specifies the elements it applies to.

Although in our examples the entities (e.g., players) only contain their name,
in practice a real application would store additional information about
each entity (e.g., personal details).
For supporting restoring information associated with entities we have
added a touch operation that acts as an add for determining if the element
is in the collection, but preserves the information that was associated
with the entity.
When an effect is added to an operation, a touch should be used instead of an add.
This operation is implemented by keeping removed elements and using SwiftCloud
stability information for garbage-collection.

%

\subsubsection{Compensation CRDTs}
\label{sec:compensations}
For some constraints, it is possible to encapsulate the logic for detecting conflicts and
applying the compensations automatically, reducing the effort for checking if some
constraint holds and applying compensations across the application.
For example, consider the constraint that enforces a maximum number of players in
a tournament.
We use a set to store the information about enrolled players in a tournament.
To ensure that the application is always consistent, whenever the application
accesses the set,
the code would have to check if the size is within limits, and, in case it was not,
it would have to apply the defined compensation.

Our Compensations Set CRDT allows the programmer to define the constraint that
must be maintained at all times, and the compensation that must execute, when
it is false.
Whenever the object is read, the code is executed automatically, ensuring that
any observed state is consistent.
The effects of the compensation, in case it is executed, are committed alongside with
the effects of the operation that accessed the customized set.

In case a compensation has to remove some element from the set, the element
is chosen deterministically.
This does not guarantee that more elements than necessary are removed, but it reduces
the chance of that happening.
As the operation is propagated to all replicas, it guarantees that if an element is
removed at one site, then it is
removed at all sites, ensuring convergence.

\section{Evaluation}
\label{sec:eval}

In this section, we present an evaluation of \name{}, meant to answer the following
questions:
\begin{asparaenum}[(i)]
    \item Which invariants are covered by our approach?
    \item What is the effort of using \name{}?
	\item How does the performance of applications modified by \name{} compare to
	other solutions that maintain invariants but use coordination in detriment of performance, or do not maintain
	invariants?
\end{asparaenum}

\subsection{Invariant preservation with \name{}}
\label{sec:eval-part-one}

This section surveys the invariants covered by our approach
by analyzing the use of \name{} in several applications.

\begin{table}[t]
\center
{\footnotesize
\renewcommand\tabcolsep{5pt}
\begin{tabular}{c | c c | c c c c}
Inv. Type        & \textit{I}-Conf. & \name{} & TPC & Tour & Ticket & Twitter\\
\hline
Sequential id. & No        & No         & Yes     & ---        & --- &  ---\\
Unique id.     & Yes       & Yes        & Yes     & Yes        & Yes & Yes\\
Numeric inv.    & No & Comp.      & Yes     & ---        &  --- &  ---\\
Aggreg. const. & No & Comp.    & ---     & ---        & --- &  ---\\
Aggreg. incl.  & Yes  & Yes        & ---     & Yes        & --- &  ---\\
Ref. integrity & No   & Yes        & Yes     & Yes        & --- &  Yes\\
Disjunctions   & No  & Yes        & ---     & Yes        & --- &  ---\\
\end{tabular}
}
\caption{Types of Invariants present in applications.}
\label{table:invariants}
\end{table}

\subsubsection{Classes of invariants}
\label{sec:eval:invariantclasses}

Prior work has analyzed the invariants that are used in real applications~\cite{feralcc,bailisiser,redblue}.
Table~\ref{table:invariants} summarizes whether they can be preserved using \WC{}
only \cite{bailisiser} (\emph{I-Confluent}) or using \emph{\name{}}.

\textbf{Sequential identifiers:} Sequential identifiers are useful for
enforcing an ordering of elements.
In general, generating these identifiers requires coordination to avoid collisions.
No solution, based on weak consistency can maintain this invariant.
However, it has been shown that, in most cases, applications could easily replace
the use of sequential identifiers by unique identifiers~\cite{hat,speedy}.

\textbf{Unique identifiers:} Unique identifiers can be preserved without coordination at runtime. It suffices to pre-partition the space of identifiers among the nodes that will generate them.

\textbf{Numeric invariants:} Numeric invariants assert conditions
involving numeric predicates (e.g., \emph{$p < k$}).
In general, preserving these invariants requires coordination. However, support is possible on top of weak consistency by relying on escrow techniques~\cite{bcounters,disciplined-inconsistencies,ONeil1986Escrow}.
In \name{}, we can use compensations to preserve this type of invariants,
whenever the semantics is reasonable for the application~\cite{cap12years}.
For example, to replenish the stock of a product, like in TPC-C/W.

\textbf{Aggregation constraint:} Imposing a bound on the size of a collection,
e.g., limiting the players enrolled in a tournament, can be addressed using a numeric
invariant over a predicate that represents the size of the collection, thus sharing
its properties.

\textbf{Aggregation inclusion:} Ensuring an element is eventually added
or removed from a collection is \IConfluent{}, provided no dependencies to other objects exist. If that is not the case,  then preserving referential integrity is usually required.

\textbf{Referential integrity:} Preserving relations and dependencies among objects, such as foreign keys in relational databases and references to keys in key-value stores, is not \IConfluent{}. \name{} fully supports this invariant, as exemplified throughout the paper.

\textbf{Disjunctions:}
Applications often specify that one of several conditions must be met
by using a disjunction.
\name{} can address this type of invariant by extending an operation
to ensure that the disjunction is always true.
This is an extension of the mechanism for supporting referential integrity,
as in this case there might be several alternative conditions that restore
the validity of the invariant.

\subsubsection{Invariants in applications}
\label{sec:applications}

We now analyze how \name{} can address the invariants that are present
in several representative applications (summarized in Table \ref{table:invariants}).


\textbf{\tournamentApp{}} This application showcases some of the invariants that our solution can address.
It is based on one used in prior work~\cite{cise,indigo} and includes new operations with more constraints.
For this application, \name{} is capable of proposing multiple alternative resolutions that either reconstruct
broken dependencies, or clear them, to avoid inconsistencies due to concurrent executions, as discussed
throughout this paper.
%

\textbf{\twitter{}} We implemented a clone of Twitter that relies heavily on referential integrity to implement user timelines and maintain subscribers information. When some user tweets, we opted for writing immediately to all followers timelines. This emphasises consistency issues that arise when tweets or users are removed concurrently. Our version explores several alternatives for solving these conflicts. If a tweet is retweeted and removed concurrently, the options are to recover the deleted tweet or hide all of its retweets from the followers timelines.
As for handling user removals, \name{} can leverage the \remwins{} semantics to purge all the user's history from the timelines of the other users concurrently with any other operations that might be happening.


\textbf{\ticket{}:} this application is based on FusionTicket~\cite{fusionticket,disciplined-inconsistencies,salt-osdi14}.
The main invariant of this application is that tickets for events cannot be oversold.
It is necessary to use compensations in this case, as it is impossible to prevent
the violation beforehand, as discussed in~\ref{sec:extensions}.
When the tickets available are oversold, the application cancels the ticket and
reimburses the user. The transfer of money to the client's account crosses the boundaries of the
system thus it must use a different mechanism.

\textbf{\tpcw{} and \tpcc{}:} These
standard database benchmarks overlook some aspects of
real-world applications, such as having operations to manage product listings.
In our specification, we extended these applications to include such operations, which introduced referential integrity constraints. For addressing the lack of inventory after purchase, we used \name{} compensations to increase the stock (as in the specification of the benchmark). An alternative would be to cancel the
oversold purchases, as in the previous example.



\subsubsection{Using the \name{} tool}
\label{sec:applications:modified}



The \name{} algorithm generates new operation specifications
by testing conflicts and augmenting operations, in an iterative process, supervised by the programmer.
The number of tests that our tool generates is bounded by the number of operations
in the application specification. We rely on a SMT solver~\cite{de2008z3} to test all valid combinations of
parameters efficiently. Despite the satisfiability problem having exponential complexity, the solver is capable
of handling most cases in polynomial time. In our tests, using a modern laptop, this automatic step of the algorithm was fast enough to not hinder interactivity and frustrate the programmer.

\begin{figure}[t]
	\hspace{0.2cm}
	\scalebox{.82}{		\includecode{ensure-new.java}	}
\caption{Auxiliary functions that restore preconditions in the \tournamentApp{} application.}
\label{lst:ensure}
\end{figure}

In terms of the work required to write the modified version of the application, this
effort is small.
For example, Figure~\ref{lst:ensure} presents the code of the auxiliary functions
that are necessary to restore the consistency of the \tournamentApp{} application.
The other applications that we have implemented follow a similar scheme.
Only a few lines of code are necessary to add to each conflicting operation.

\subsubsection{Discussion}
The invariants that the \name{} tool can support are limited to the extent of
invariants that can be expressed using the language that we have defined.
The classes of invariants that we support (Figure~\ref{table:invariants}) are
common in many internet application, as remarked by Bailis et al.~\cite{feralcc}.
The examples discussed in the previous section show that the
language is expressive enough to address rather complex applications,
including typical relational database applications.

If a database is shared by multiple applications, the programmer must
create a single specification of all applications for the analysis to identify
all possible conflicts.
The alternative would be to provide the resolution mechanisms at the
storage level and to repair invariants independently of the applications
developed on top.
We chose to apply transformations at application level to show the possibility
of implementing the applications without changing the underlying storage.

The effort of writing specifications is arguably comparable to the effort of
writing the code itself~\cite{Parnas2011}.
A lot of research has dealt with this problem, proposing automatic feature
extraction, and code synthesis, to aid the programmer in writing correct
applications~\cite{Roy14Writes,Li14Automating,Ernst200735,Flanagan2001,blazes}.
Our approach stands to benefit from these complementary research avenues.

\subsection{Performance evaluation}
\label{sec:eval-part-two}

In this section, we compare the performance of modified applications against other solutions.
We expect the modifications to have a minimal overhead
in comparison to the original code running on \WC{}. We also expect the latency
of the operations to be clearly lower in comparison to systems that use coordination
to enforce invariant preservation.
We also try to measure the tipping point at which solutions based on coordination
are faster than executing extra updates. For this, we use synthetic benchmarks.

\subsubsection{System configurations}
The benchmarks execute in a geo-replicated setting on Amazon EC2.
The database deployment consists of three servers running in three geographical
regions, with mean latency around 80 milliseconds between US-EAST and US-WEST
and US-EAST and EU-WEST, and 160 between EU-WEST and US-WEST.

The application server is co-located with the storage system of each region.
We use SwiftCloud to implement all different approaches that we evaluate.
Clients are installed in other machines in the same availability zones
as the corresponding closest servers.

We compare the performance of applications with the following configurations:

\noindent{\textbf{Causal Consistency (\Causal{})}} Unmodified applications, does not maintain invariants for conflicting operations.

\noindent{\textbf{Inv. Preserving Applications (\name{})}} Applications modified using \name{},
maintains
invariants on top of \emph{Causal}.

\noindent{\textbf{Strong Consistency (\Strong{})} all update operations are forwarded to
a single server to enforce serialization. We use the US-EAST replica to execute
updates and to minimize the average latency.

\noindent{\textbf{Invariant violation avoidance (\Indigo{}~\cite{indigo})}}
Applications modified with coordination mechanisms to prevent
conflicting operations from executing concurrently. We use Indigo\footnote{Indigo is implemented on top of SwiftCloud. The code was made available by the authors at \url{https://github.com/SyncFree/Indigo}.} for implementing this configuration.
In Indigo, a conflicting
operation needs to possess or acquire the reservations needed for safe execution under concurrency.
Reservations can be exchanged and shared between
replicas asynchronously in a pairwise fashion, which is usually cheaper than
full coordination among all replicas.

\subsubsection{Throughput and latency}

We evaluate the scalability of each configuration of the system by measuring
the latency of operations with different loads on the system, using the \tournamentApp{}
application.
The workload comprises $35\%$ of write operations.
All operations are conflicting in the original specification.
In the modified version, \name{}, all operations are \IConfluent{}
and use a mix of conflict resolution policies.
In the Indigo implementation, every pair of operations is protected
by a reservation.

To test the scalability of the system, we increase the number of clients contacting
each server by running extra client threads until peak throughput is achieved.

The results show that \Strong{} presents the highest average latency,
which is a consequence of having $\frac{2}{3}$ of operations being forwarded to a remote server.
\Causal{} shows the best scalability
with the lowest latency. Our approach, \name{}, performs slightly
worse than \Causal{}, as additional updates need to be executed,
but enforces application invariants.
When compared to \Indigo{}, our approach performs slightly better.
The advantage is small because, while each operation requires acquiring a reservation, reservations
are exchanged among replicas very infrequently after that.

Figure~\ref{fig:tour-lat-bars} presents the latency for the write
operation and highlights more clearly the differences between
the configurations.
We omit the \SC{} column.
The average latency of operations in \Indigo{} is higher than the latency for
\Repair{} or \Causal{} and also exhibits a greater standard deviation. Both are explained
by the occasional need for \Indigo{} replicas to trade reservations.
Compared to \Causal{}, the latency of the write operations is only slightly higher in the~\name{} approach, which is due to the extra code they execute.
The overhead of executing extra effects is discussed in Section~\ref{sec:mic-bench}.

\subsubsection{Comparing different strategies}
We implemented \twitter{} using \addwins{} and
\remwins{} strategies to compare the overheads of each
approach.
Figure~\ref{fig:twitter-lat-bars} shows the latency of each
operation for the different strategies.
The \addwins{} version must ensure that when
a user tweets, or retweets, he cannot be removed
concurrently. This incurs in the cost of restoring the user for those operations and explains their higher latency compared to \Causal{}.
Whereas, \remwins{} strategy must ensure that no user reads a tweet that was removed concurrently.
Pessimistically, this would have to remove the tweet from the timelines of every user in the system, as the tweet
could be added to anyone's timeline, concurrently. Instead, we enact this strategy with a compensation, applied when accessing user timelines. This hides tweets that were removed, thus restoring the invariant, trading a slightly higher latency in reads to prevent unnecessary writes.


\begin{figure*}[t]
    \centering
    \begin{minipage}{0.31\textwidth}
		\includegraphics[width=\textwidth]{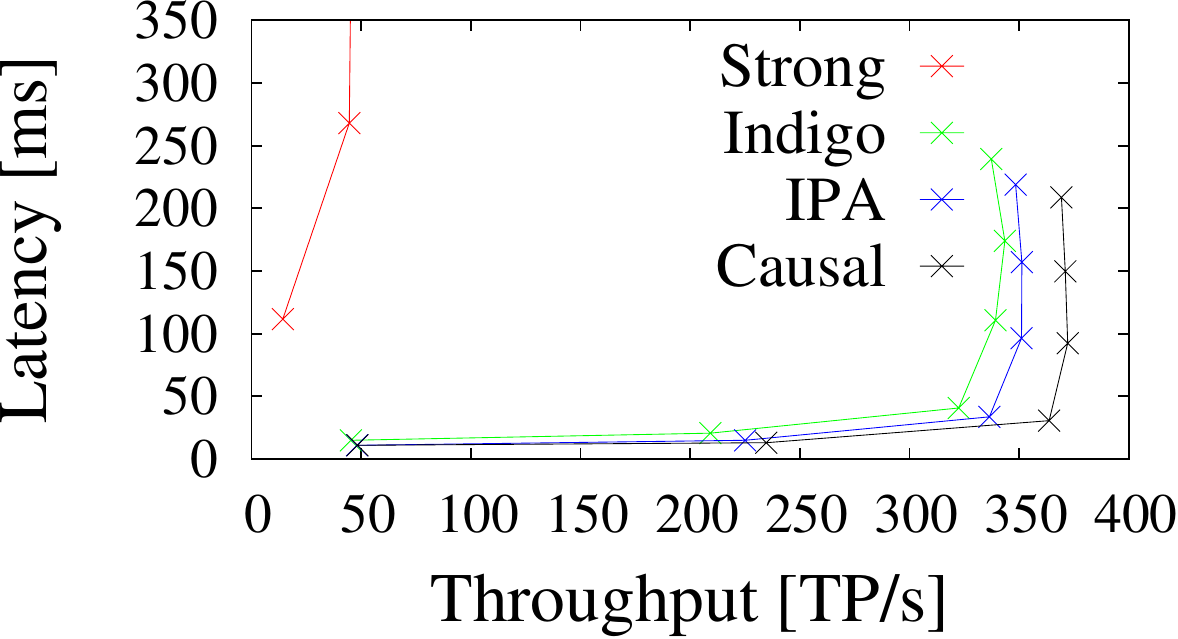}
    	\caption{Peak throughput for \tournamentApp{}.}
	    \label{fig:tourn-tp-lat}
	\end{minipage}
	\hspace{0.2cm}
    \begin{minipage}{0.32\textwidth}
		\includegraphics[width=\textwidth]{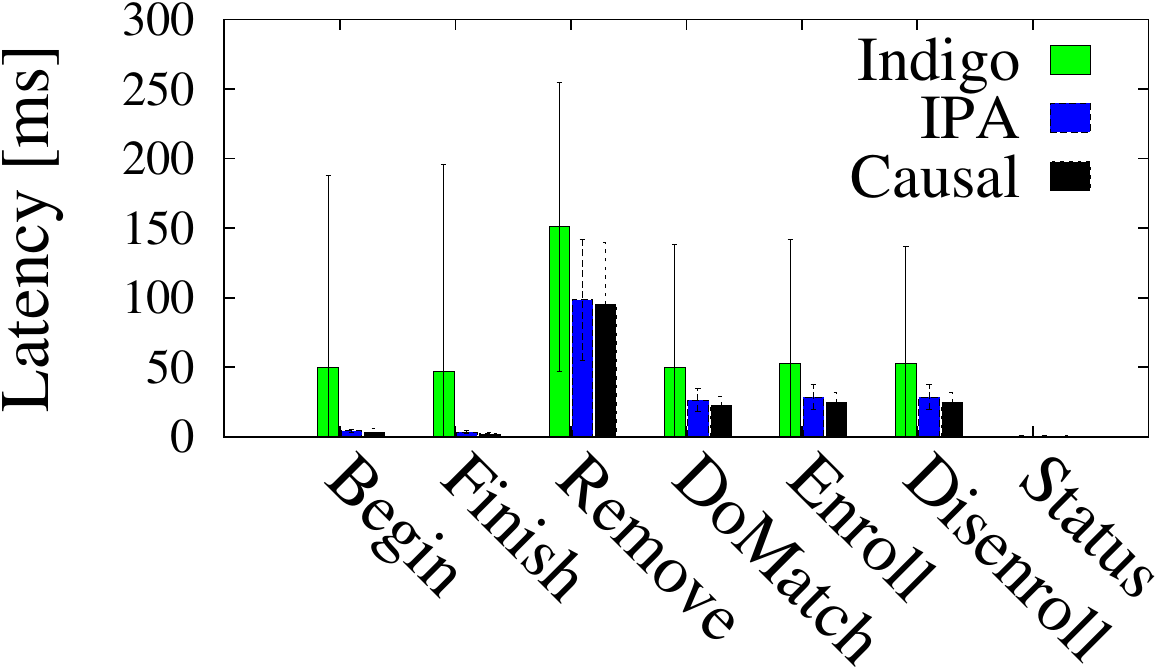}
    	\caption{Latency of individual operations in \tournamentApp{}.}
	    \label{fig:tour-lat-bars}
	\end{minipage}
	\hspace{0.2cm}
    \begin{minipage}{0.32\textwidth}
		\includegraphics[width=\columnwidth]{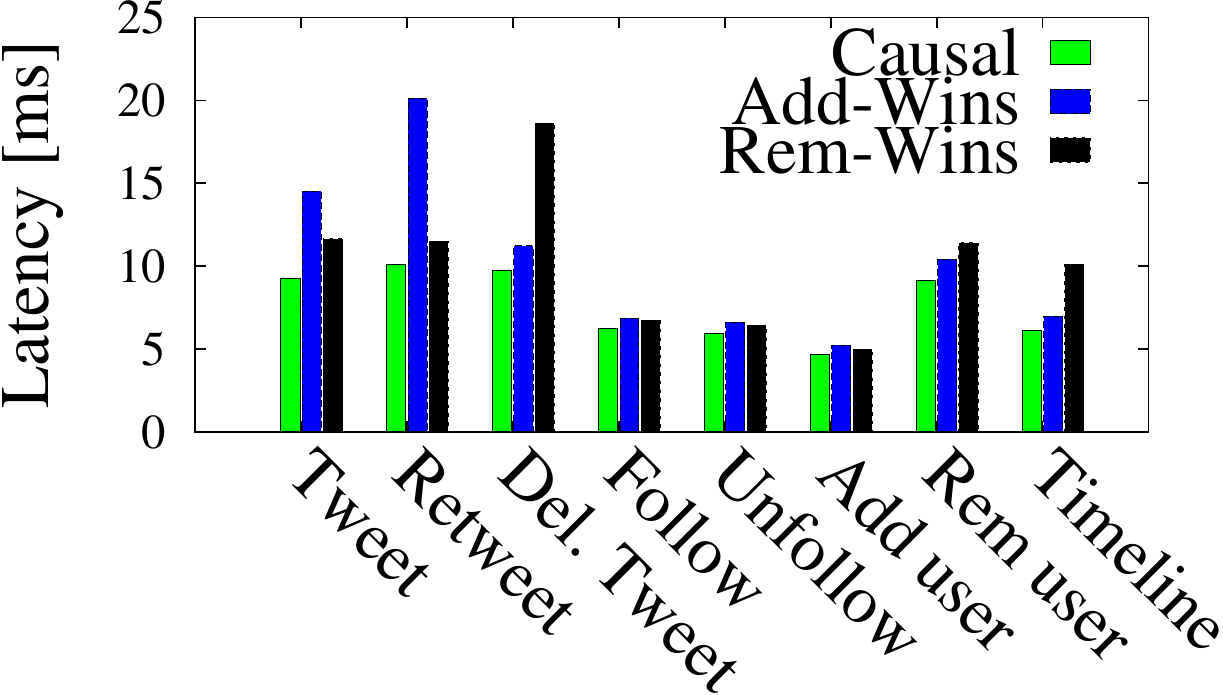}
		\caption{Latency of individual operations in \twitter{}.}
	    \label{fig:twitter-lat-bars}
	\end{minipage}
         \vspace{-6px}
\end{figure*}

\subsubsection{Compensations scalability}
We evaluate the scalability of the compensations CRDT
in the \ticket{} application, by increasing contention.
Figure~\ref{fig:ticket-tp-lat} presents the latency of operations
for a certain load of the server.
The red dots in the figure indicate the average number of invariant violations
that were observed at that throughput, when using \Causal{}.
They confirm the intuition that as contention rises, the divergence
window grows larger, increasing the chance for invariant violation.
In \Causal{}, this exposes the application consistency anomalies, while
in \name{} executing the compensations preserves the invariants at all times.
As expected, compensations incur on some overhead,
but still provide latency comparable to \Causal{}.

\subsubsection{Microbenchmarks}
\label{sec:mic-bench}

\name{} avoids invariant violations by adding extra updates to one
or multiple objects.
In this section, we evaluate the overhead of adding additional effects to operations.
We analyze the impact of executing increasingly more updates in comparison to the
costs of executing the original operation in \SC{} and Indigo.

\textbf{Operations on a single object:}
We measure the speedup of an application running on top
of causal consistency that executes extra updates for a single object versus
the original operation running on \Strong{}.
Figure~\ref{fig:micro-over-keys} (top) shows that the original operation is
about $28 \times$ faster in \Repair{} than in \Strong{}.
Adding more updates to this operation makes the speedup decrease.
When we execute 2048 updates to a single object, the average latency is
still about 40ms.

\textbf{Operations over multiple objects:}
Executing updates on a single object imposes a low overhead on the system,
because the object is read and written to storage only once and subsequent updates
only impose processing costs.
Now we evaluate the overhead when modified operations have updates over multiple objects.

The original application reads a varying number of objects
to check some condition and then executes a single write operation
to an object.
The modified application checks the same condition, but executes a write for each
object.
Figure~\ref{fig:micro-over-keys} (bottom) shows performance dropping
faster than when executing updates over single objects.
At 64 objects, it starts to pay off to switch to \Strong{}.

In practice, in the applications that we evaluated, we require only a few extra updates
per object over a small number of objects. In the case of \twitter{}, which needs to execute more writes due to our implementation of the timeline, we were able to execute them lazily via compensations.

\begin{figure*}[t]
    \centering
    \begin{minipage}{0.31\textwidth}
		\includegraphics[width=\textwidth]{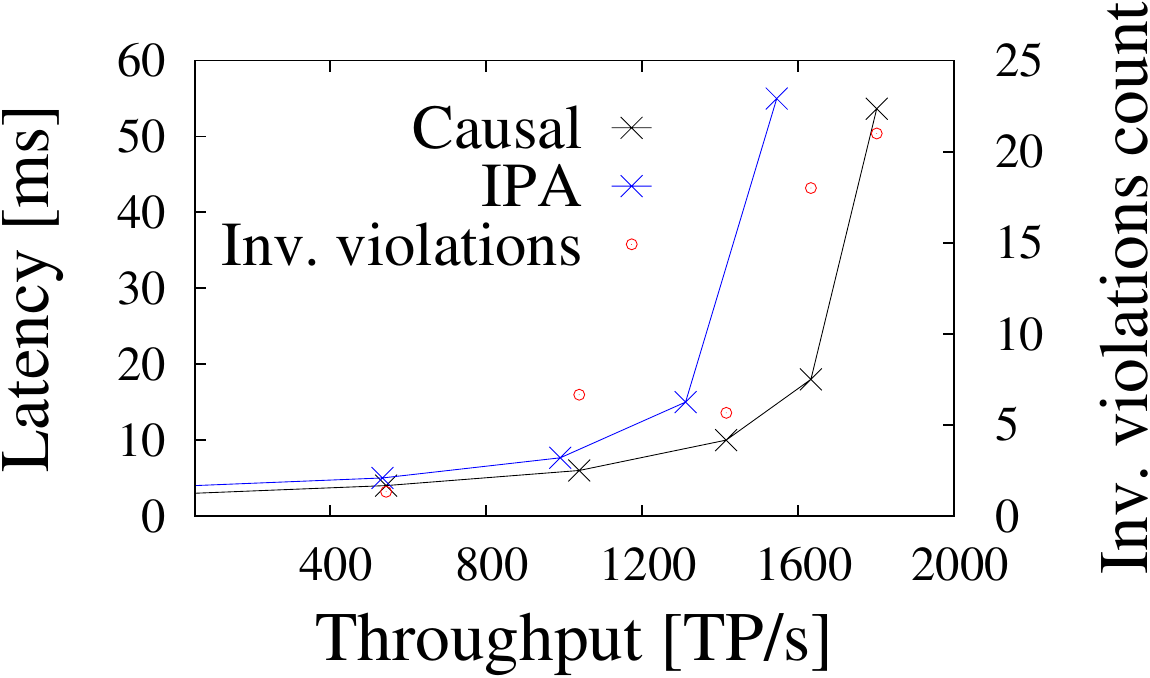}
    	\caption{Peak throughput for \ticket{} benchmark. Red dots indicate number of invariant violations observed during runtime.}
	    \label{fig:ticket-tp-lat}
	\end{minipage}
	\hspace{0.2cm}
    \begin{minipage}{0.26\textwidth}
        \includegraphics[width=\columnwidth]{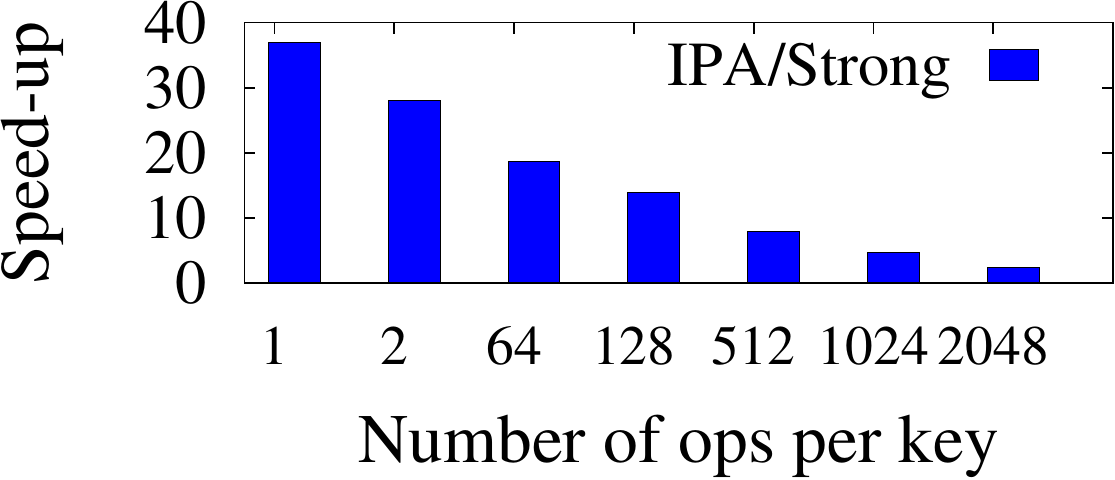}
    	\label{fig:micro-over-ops}
        \includegraphics[width=\columnwidth]{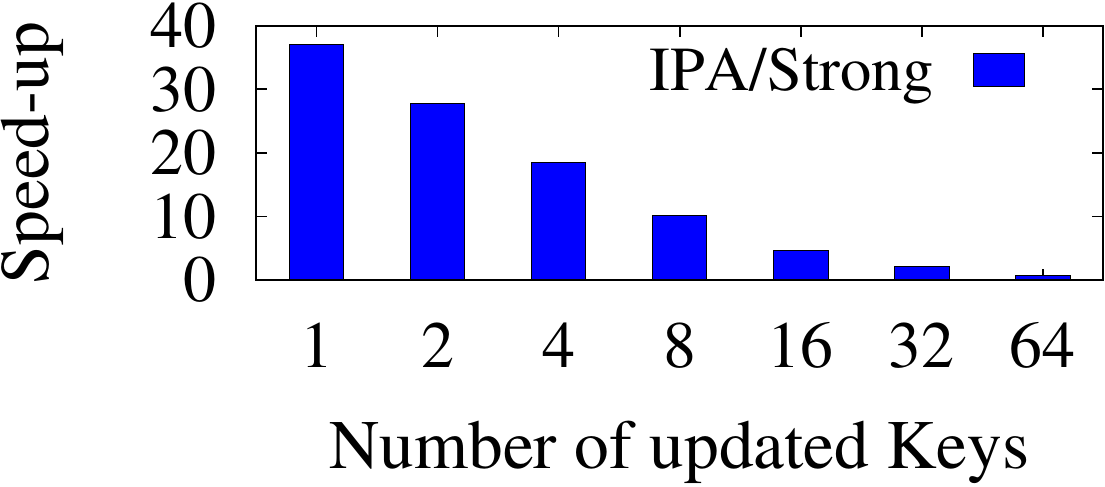}
    	\caption{Speed-up of executing multiple writes in IPA versus \Strong{}.}
    	\label{fig:micro-over-keys}
	\end{minipage}
	\hspace{0.2cm}
    \begin{minipage}{0.31\textwidth}
		\includegraphics[width=\columnwidth]{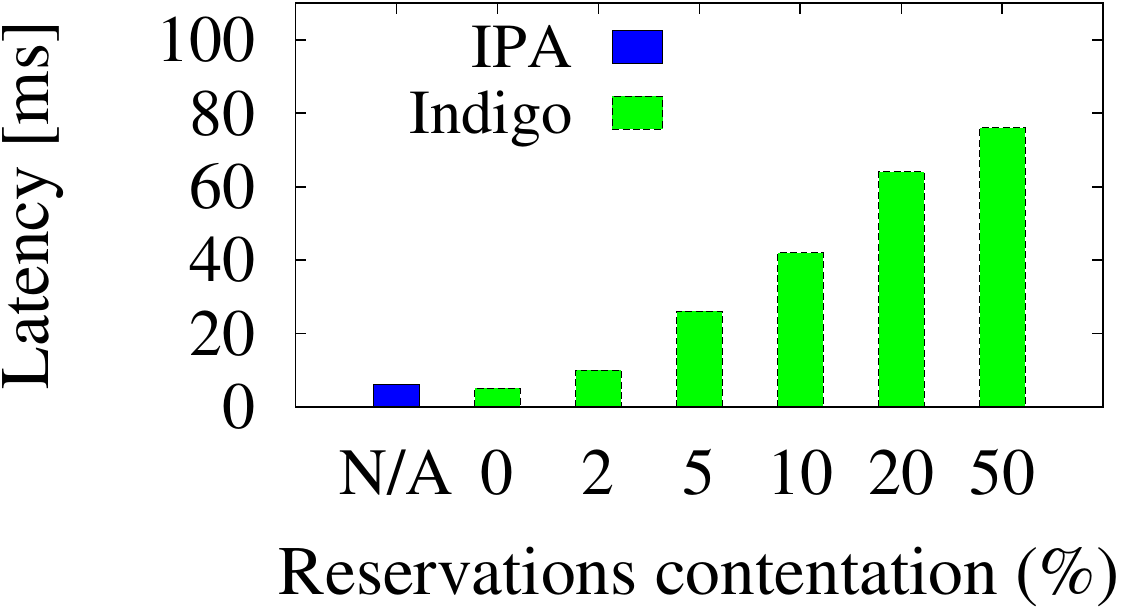}
	    \caption{Latency of operations with varying percentage of reservation types in comparison to no reservations.}
		\label{fig:micro-over-indigo}
	\end{minipage}
         \vspace{-6px}
\end{figure*}

\textbf{Comparison with \Indigo{}:}
In Indigo \cite{indigo}, operations are allowed to execute locally if the replica holds some
specific reservation.
Multiple operations might be able to execute concurrently at different replicas
if all of them can share the same reservation.
If a replica requires some reservation that is being used exclusively, it must request remote replicas
to release it, before acquiring it.
This approach only avoids coordination when a replica holds the necessary reservations to
execute some operation. Thus, the latency of an application depends on the contention
for obtaining the reservations.

In this experiment, we evaluate the impact of varying the percentage of
operations that compete to acquire some reservations.
We compare the performance of this solution against executing the same operation in \name{}.
Figure~\ref{fig:micro-over-indigo} shows that \Repair{} performance
is equivalent to \Indigo{} with no contention for reservations, and that the latency
of \Indigo{} rises steadily as contention increases.

Despite the overhead for executing the additional effects, \name{}
provides a predictable latency for operations, which is not the case
for \Indigo{}, whose operations latency depend on the current distribution
of reservations. Furthermore, our approach is fault-tolerant as a client can execute
operations as long as it can access a single server. In \Indigo{}, if a server
that holds the necessary reservation to execute some operation becomes unavailable,
the operation cannot be executed.

\section{Related work}
\label{sec:relatedwork}

Achieving low latency, high availability and data consistency in distributed systems
is difficult, as postulated in the CAP theorem~\cite{cap}.
In recent years, researchers and practitioners have studied the trade-offs
in distributed systems to provide the best consistency guarantees for different types of
applications~\cite{spanner,dynamo,cops,hat,redblue,indigo,transactionchains}.

Systems that ensure strong isolation criteria~\cite{spanner,transactionchains,farm,mdcc} require
coordination across replicas, which is expensive in geo-replicated scenarios.
In Megastore~\cite{megastore}, data is partitioned at a fine granularity to achieve low latency,
while MDCC~\cite{mdcc} exploits commutativity and protocol optimizations to improve
performance.
Spanner~\cite{spanner} and Farm~\cite{farm} harness custom hardware to improve
performance~\cite{spanner,farm}.

Systems that use weak consistency are widely deployed in the real-world
\cite{dynamodb,cassandra,riak,bashoclients,cassandraclients}, but can be difficult to
use~\cite{feralcc}.
Many systems provide causal consistency coupled with object
convergence and transactions~\cite{cops,swiftcloud,eiger,hat}, which all can be implemented
efficiently without hindering the availability of the system.

Convergent data types~\cite{crdts} provide automatic replica convergence,
which lessens the programming effort in these systems.
However, data type convergence alone cannot prevent invariant violations involving multiple objects.
To mitigate the problem, RedBlue~\cite{redblue} and Walter~\cite{walter} provide support
for executing operations under weak or strong consistency to allow fast operations when
invariants are not at risk, and consistent operations otherwise~\cite{redblue,walter}.
Sieve~\cite{Li14Automating} and Blazes~\cite{blazes} address the problem of automating
the use of the most appropriate consistency alternative, while Indigo~\cite{indigo} and the
Homeostasis protocol~\cite{Roy14Writes}
try to minimize the use of the strong consistency path. Despite improving the latency of
operations in the general case, systems that depend on coordination to execute some
operations may still become unavailable and exhibit high latency.

Helland and Campbell have suggested that applications should handle invariant
violations as part of the application  logic, as an alternative to executing operations under strong consistency to prevent violations~\cite{quicksand}.
The idea of compensations~\cite{sagas} is to execute operations optimistically and explicitly
rollback the effects when conflicts are detected.
A few systems have explored this model.
In PLANET~\cite{planet}, transactions execute speculatively, allowing the system to
provide the control back to the client before the transaction commit confirmation arrives.
In Bayou~\cite{bayou}, transactions commit locally and remain in a tentative state, until all
replicas  agree on the ordering of operations.
Existing systems that use compensations still use some form of coordination to
commit transactions.
Our approach departs from this model by modifying the operations to ensure they can always commit locally while preserving invariants.
We show that our approach does not alter the semantics of operations
when no conflicting concurrent operations execute.

A recent line of work focuses on proving correctness of distributed
systems \cite{ironfleet,verdi}.
These proposals complement our own, since their focus is on attesting
if implementations conform to some specification,
whereas we aim to provide a methodology for implementing correct applications on top of the assumptions of our chosen consistency model.

\section{Conclusion}
\label{sec:conclusion}

In this paper, we propose a novel approach for supporting correct, highly available
applications on top of weak consistency, based on the definition of
correctness criteria for each application.
A static analysis step identifies which operations can lead to invariant violation
when executed concurrently, and augments them
with additional effects to enable state convergence and invariant preservation
without the need for coordination.
We show that our approach extends the range of invariants that can be made \IConfluent{},
while preserving sensible operation semantics.

Experimental results back the viability of the approach, showing that
the modified applications present a performance similar to the original applications.
The features required from the underlying storage system
are available in several existing weakly consistent databases, which can ease
the adoption of the proposed approach in real-life applications.



\bibliographystyle{abbrv}
\bibliography{bib}

\begin{thebibliography}{10}

\bibitem{bashoclients}
{Companies Using NoSQL}.
\newblock \url{http://basho.com/about/customers/}.
\newblock Accessed May-2016.

\bibitem{cassandraclients}
{Companies Using NoSQL Apache Cassandra}.
\newblock \url{http://www.planetcassandra.org/companies/}.
\newblock Accessed May-2016.

\bibitem{fusionticket}
{Fusion Ticket}.
\newblock \url{http://www.fusionticket.org/}.
\newblock Accessed September-2016.

\bibitem{antidote}
D.~D. Akkoorath, A.~Z. Tomsic, M.~Bravo, Z.~Li, T.~Crain, A.~Bieniusa,
  N.~Preguiça, and M.~Shapiro.
\newblock {Cure: Strong semantics meets high availability and low latency}.
\newblock In {\em Proc. of the 36th IEEE International Conference on
  Distributed Computing Systems (ICDCS 2016)}, Nara, Japan, June 2016.

\bibitem{blazes}
P.~Alvaro, N.~Conway, J.~M. Hellerstein, and D.~Maier.
\newblock Blazes: Coordination analysis for distributed programs.
\newblock In {\em {IEEE} 30th International Conference on Data Engineering,
  Chicago, {ICDE} 2014, IL, USA, March 31 - April 4, 2014}, pages 52--63, 2014.

\bibitem{hat}
P.~Bailis, A.~Davidson, A.~Fekete, A.~Ghodsi, J.~M. Hellerstein, and I.~Stoica.
\newblock Highly available transactions: Virtues and limitations.
\newblock {\em Proc. VLDB Endow.}, 7(3):181--192, Nov. 2013.

\bibitem{bailisiser}
P.~Bailis, A.~Fekete, M.~J. Franklin, A.~Ghodsi, J.~M. Hellerstein, and
  I.~Stoica.
\newblock Coordination avoidance in database systems.
\newblock {\em Proc. VLDB Endow.}, 8(3):185--196, Nov. 2014.

\bibitem{feralcc}
P.~Bailis, A.~Fekete, M.~J. Franklin, A.~Ghodsi, J.~M. Hellerstein, and
  I.~Stoica.
\newblock Feral concurrency control: An empirical investigation of modern
  application integrity.
\newblock In {\em Proceedings of the 2015 ACM SIGMOD International Conference
  on Management of Data}, SIGMOD '15, pages 1327--1342, New York, NY, USA,
  2015. ACM.

\bibitem{megastore}
J.~Baker, C.~Bond, J.~C. Corbett, J.~Furman, A.~Khorlin, J.~Larson, J.-M. Leon,
  Y.~Li, A.~Lloyd, and V.~Yushprakh.
\newblock Megastore: Providing scalable, highly available storage for
  interactive services.
\newblock In {\em Proceedings of the Conference on Innovative Data system
  Research (CIDR)}, pages 223--234, 2011.

\bibitem{indigo}
V.~Balegas, S.~Duarte, C.~Ferreira, R.~Rodrigues, N.~Pregui\c{c}a,
  M.~Najafzadeh, and M.~Shapiro.
\newblock Putting consistency back into eventual consistency.
\newblock In {\em Proceedings of the Tenth European Conference on Computer
  Systems}, EuroSys '15, pages 6:1--6:16, New York, NY, USA, 2015. ACM.

\bibitem{bcounters}
V.~Balegas, D.~Serra, S.~Duarte, C.~Ferreira, M.~Shapiro, R.~Rodrigues, and
  N.~Preguiça.
\newblock Extending eventually consistent cloud databases for enforcing numeric
  invariants.
\newblock In {\em Reliable Distributed Systems (SRDS), 2015 IEEE 34th Symposium
  on}, pages 31--36, Sept 2015.

\bibitem{riak}
Basho.
\newblock Using data types -- riak documentation.
\newblock \url{http://docs.basho.com/riak/latest/dev/using/data-types/}.
\newblock Accessed Feb-2016.

\bibitem{cap12years}
E.~Brewer.
\newblock Cap twelve years later: How the "rules" have changed.
\newblock {\em Computer}, 45(2):23--29, Feb 2012.

\bibitem{pnuts}
B.~F. Cooper, R.~Ramakrishnan, U.~Srivastava, A.~Silberstein, P.~Bohannon,
  H.-A. Jacobsen, N.~Puz, D.~Weaver, and R.~Yerneni.
\newblock {PNUTS: Yahoo!'s Hosted Data Serving Platform}.
\newblock {\em Proc. VLDB Endow.}, 1(2):1277--1288, Aug. 2008.

\bibitem{spanner}
J.~C. Corbett, J.~Dean, M.~Epstein, A.~Fikes, C.~Frost, J.~J. Furman,
  S.~Ghemawat, A.~Gubarev, C.~Heiser, P.~Hochschild, W.~Hsieh, S.~Kanthak,
  E.~Kogan, H.~Li, A.~Lloyd, S.~Melnik, D.~Mwaura, D.~Nagle, S.~Quinlan,
  R.~Rao, L.~Rolig, Y.~Saito, M.~Szymaniak, C.~Taylor, R.~Wang, and
  D.~Woodford.
\newblock {Spanner: Google's Globally-distributed Database}.
\newblock In {\em Proc.\ 10th USENIX Conf.\ on Operating Systems Design and
  Implementation}, OSDI'12, pages 251--264, Berkeley, CA, USA, 2012. USENIX
  Association.

\bibitem{z3}
L.~De~Moura and N.~Bj{\o}rner.
\newblock {Z}3: An efficient {SMT} solver.
\newblock In {\em Tools and Algorithms for the Construction and Analysis of
  Systems}, TACAS '08, pages 337--340. Springer, 2008.

\bibitem{de2008z3}
L.~De~Moura and N.~Bj{\o}rner.
\newblock {Z}3: An efficient {SMT} solver.
\newblock In {\em Tools and Algorithms for the Construction and Analysis of
  Systems}, TACAS '08, pages 337--340. Springer, 2008.

\bibitem{dynamo}
G.~DeCandia, D.~Hastorun, M.~Jampani, G.~Kakulapati, A.~Lakshman, A.~Pilchin,
  S.~Sivasubramanian, P.~Vosshall, and W.~Vogels.
\newblock {Dynamo: Amazon's Highly Available Key-value Store}.
\newblock In {\em Proc.\ 21st ACM SIGOPS Symp.\ on Operating Systems
  Principles}, SOSP '07, pages 205--220, New York, NY, USA, 2007. ACM.

\bibitem{farm}
A.~Dragojevi\'{c}, D.~Narayanan, E.~B. Nightingale, M.~Renzelmann, A.~Shamis,
  A.~Badam, and M.~Castro.
\newblock No compromises: Distributed transactions with consistency,
  availability, and performance.
\newblock In {\em Proceedings of the 25th Symposium on Operating Systems
  Principles}, SOSP '15, pages 54--70, New York, NY, USA, 2015. ACM.

\bibitem{Ernst200735}
M.~D. Ernst, J.~H. Perkins, P.~J. Guo, S.~McCamant, C.~Pacheco, M.~S. Tschantz,
  and C.~Xiao.
\newblock The daikon system for dynamic detection of likely invariants.
\newblock {\em Science of Computer Programming}, 69(1–3):35 -- 45, 2007.
\newblock Special issue on Experimental Software and Toolkits.

\bibitem{Flanagan2001}
C.~Flanagan and K.~R.~M. Leino.
\newblock Houdini, an annotation assistant for esc/java.
\newblock In J.~N. Oliveira and P.~Zave, editors, {\em FME 2001: Formal Methods
  for Increasing Software Productivity: International Symposium of Formal
  Methods Europe Berlin, Germany, March 12--16, 2001 Proceedings}, pages
  500--517, Berlin, Heidelberg, 2001. Springer Berlin Heidelberg.

\bibitem{sagas}
H.~Garcia-Molina and K.~Salem.
\newblock Sagas.
\newblock In {\em Proceedings of the 1987 ACM SIGMOD International Conference
  on Management of Data}, SIGMOD '87, pages 249--259, New York, NY, USA, 1987.
  ACM.

\bibitem{cap}
S.~Gilbert and N.~Lynch.
\newblock Brewer's conjecture and the feasibility of consistent, available,
  partition-tolerant web services.
\newblock {\em SIGACT News}, 33(2):51--59, June 2002.

\bibitem{cise}
A.~Gotsman, H.~Yang, C.~Ferreira, M.~Najafzadeh, and M.~Shapiro.
\newblock {'Cause I'm Strong Enough: Reasoning About Consistency Choices in
  Distributed Systems}.
\newblock {\em SIGPLAN Not.}, 51(1):371--384, Jan. 2016.

\bibitem{ironfleet}
C.~Hawblitzel, J.~Howell, M.~Kapritsos, J.~R. Lorch, B.~Parno, M.~L. Roberts,
  S.~Setty, and B.~Zill.
\newblock Ironfleet: Proving practical distributed systems correct.
\newblock In {\em Proceedings of the 25th Symposium on Operating Systems
  Principles}, SOSP '15, pages 1--17, New York, NY, USA, 2015. ACM.

\bibitem{quicksand}
P.~Helland and D.~Campbell.
\newblock Building on quicksand.
\newblock In {\em {CIDR} 2009, Fourth Biennial Conference on Innovative Data
  Systems Research, Asilomar, CA, USA, January 4-7, 2009, Online Proceedings},
  2009.

\bibitem{disciplined-inconsistencies}
B.~Holt, J.~Bornholt, D.~Zhang, Irene R. K.~Ports, M.~Oskin, and L.~Ceze.
\newblock Disciplined inconsistency with consistency types.
\newblock In {\em Proceedings of the ACM Symposium on Cloud Computing},
  SOCC'16, 2016.

\bibitem{mdcc}
T.~Kraska, G.~Pang, M.~J. Franklin, S.~Madden, and A.~Fekete.
\newblock {MDCC: Multi-data Center Consistency}.
\newblock In {\em Proc.\ 8th ACM European Conf.\ on Computer Systems}, EuroSys
  '13, pages 113--126, New York, NY, USA, 2013. ACM.

\bibitem{cassandra}
A.~Lakshman and P.~Malik.
\newblock {Cassandra: A Decentralized Structured Storage System}.
\newblock {\em SIGOPS Oper. Syst. Rev.}, 44(2):35--40, Apr. 2010.

\bibitem{happensbefore}
L.~Lamport.
\newblock Time, clocks, and the ordering of events in a distributed system.
\newblock {\em Commun. ACM}, 21(7):558--565, July 1978.

\bibitem{Li14Automating}
C.~Li, J.~Leit{\~a}o, A.~Clement, N.~Pregui{\c c}a, R.~Rodrigues, and
  V.~Vafeiadis.
\newblock Automating the choice of consistency levels in replicated systems.
\newblock In {\em 2014 USENIX Annual Technical Conference (USENIX ATC 14)},
  pages 281--292, Philadelphia, PA, June 2014. USENIX Association.

\bibitem{redblue}
C.~Li, D.~Porto, A.~Clement, J.~Gehrke, N.~Pregui{\c c}a, and R.~Rodrigues.
\newblock Making geo-replicated systems fast as possible, consistent when
  necessary.
\newblock In {\em Presented as part of the 10th USENIX Symposium on Operating
  Systems Design and Implementation (OSDI 12)}, pages 265--278, Hollywood, CA,
  2012. USENIX.

\bibitem{cops}
W.~Lloyd, M.~J. Freedman, M.~Kaminsky, and D.~G. Andersen.
\newblock {Don't Settle for Eventual: Scalable Causal Consistency for Wide-area
  Storage with COPS}.
\newblock In {\em Proc.\ 23d ACM Symp.\ on Operating Systems Principles}, SOSP
  '11, pages 401--416, New York, NY, USA, 2011. ACM.

\bibitem{eiger}
W.~Lloyd, M.~J. Freedman, M.~Kaminsky, and D.~G. Andersen.
\newblock {Stronger Semantics for Low-latency Geo-replicated Storage}.
\newblock In {\em Proc.\ 10th USENIX Conf.\ on Networked Systems Design and
  Implementation}, {NSDI'13}, pages 313--328, Berkeley, CA, USA, 2013. USENIX
  Association.

\bibitem{ONeil1986Escrow}
P.~E. O'Neil.
\newblock The escrow transactional method.
\newblock {\em ACM Trans. Database Syst.}, 11(4):405--430, Dec. 1986.

\bibitem{ramcloud}
J.~Ousterhout, P.~Agrawal, D.~Erickson, C.~Kozyrakis, J.~Leverich,
  D.~Mazi\`{e}res, S.~Mitra, A.~Narayanan, G.~Parulkar, M.~Rosenblum, S.~M.
  Rumble, E.~Stratmann, and R.~Stutsman.
\newblock The case for ramclouds: Scalable high-performance storage entirely in
  dram.
\newblock {\em SIGOPS Oper. Syst. Rev.}, 43(4):92--105, Jan. 2010.

\bibitem{planet}
G.~Pang, T.~Kraska, M.~J. Franklin, and A.~Fekete.
\newblock Planet: Making progress with commit processing in unpredictable
  environments.
\newblock In {\em Proceedings of the 2014 ACM SIGMOD International Conference
  on Management of Data}, SIGMOD '14, pages 3--14, New York, NY, USA, 2014.
  ACM.

\bibitem{Parnas2011}
D.~L. Parnas.
\newblock Precise documentation: The key to better software.
\newblock In S.~Nanz, editor, {\em The Future of Software Engineering}, pages
  125--148, Berlin, Heidelberg, 2011. Springer Berlin Heidelberg.

\bibitem{Roy14Writes}
S.~Roy, L.~Kot, G.~Bender, B.~Ding, H.~Hojjat, C.~Koch, N.~Foster, and
  J.~Gehrke.
\newblock The homeostasis protocol: Avoiding transaction coordination through
  program analysis.
\newblock In {\em Proceedings of the 2015 {ACM} {SIGMOD} International
  Conference on Management of Data, Melbourne, Victoria, Australia, May 31 -
  June 4, 2015}, pages 1311--1326, 2015.

\bibitem{Schurman09latency}
E.~Schurman and J.~Brutlag.
\newblock {Performance Related Changes and their User Impact. {P}resented at
  Velocity Web Performance and Operations Conference}, 2009.

\bibitem{crdts}
M.~Shapiro, N.~Pregui\c{c}a, C.~Baquero, and M.~Zawirski.
\newblock {Conflict-free Replicated Data Types}.
\newblock In {\em Proc.\ 13th Int.\ Conf.\ on Stabilization, Safety, and
  Security of Distributed Systems}, SSS'11, pages 386--400, Berlin, Heidelberg,
  2011. Springer-Verlag.

\bibitem{dynamodb}
S.~Sivasubramanian.
\newblock Amazon dynamodb: A seamlessly scalable non-relational database
  service.
\newblock In {\em Proceedings of the 2012 ACM SIGMOD International Conference
  on Management of Data}, SIGMOD '12, pages 729--730, New York, NY, USA, 2012.
  ACM.

\bibitem{walter}
Y.~Sovran, R.~Power, M.~K. Aguilera, and J.~Li.
\newblock {Transactional Storage for Geo-replicated Systems}.
\newblock In {\em Proc.\ 23d ACM Symp.\ on Operating Systems Principles}, SOSP
  '11, pages 385--400, New York, NY, USA, 2011. ACM.

\bibitem{bayou}
D.~B. Terry, M.~M. Theimer, K.~Petersen, A.~J. Demers, M.~J. Spreitzer, and
  C.~H. Hauser.
\newblock {Managing Update Conflicts in Bayou, a Weakly Connected Replicated
  Storage System}.
\newblock In {\em Proc.\ 15th ACM Symp.\ on Operating Systems Principles}, SOSP
  '95, pages 172--182, New York, NY, USA, 1995. ACM.

\bibitem{speedy}
S.~Tu, W.~Zheng, E.~Kohler, B.~Liskov, and S.~Madden.
\newblock Speedy transactions in multicore in-memory databases.
\newblock In {\em Proceedings of the Twenty-Fourth ACM Symposium on Operating
  Systems Principles}, SOSP '13, pages 18--32, New York, NY, USA, 2013. ACM.

\bibitem{verdi}
J.~R. Wilcox, D.~Woos, P.~Panchekha, Z.~Tatlock, X.~Wang, M.~D. Ernst, and
  T.~Anderson.
\newblock Verdi: A framework for implementing and formally verifying
  distributed systems.
\newblock In {\em Proceedings of the 36th ACM SIGPLAN Conference on Programming
  Language Design and Implementation}, PLDI '15, pages 357--368, New York, NY,
  USA, 2015. ACM.

\bibitem{salt-osdi14}
C.~Xie, C.~Su, M.~Kapritsos, Y.~Wang, N.~Yaghmazadeh, L.~Alvisi, and
  P.~Mahajan.
\newblock Salt: Combining acid and base in a distributed database.
\newblock In {\em Proceedings of the 11th USENIX Conference on Operating
  Systems Design and Implementation}, OSDI'14, pages 495--509, Berkeley, CA,
  USA, 2014. USENIX Association.

\bibitem{swiftcloud}
M.~Zawirski, N.~Pregui\c{c}a, S.~Duarte, A.~Bieniusa, V.~Balegas, and
  M.~Shapiro.
\newblock Write fast, read in the past: Causal consistency for client-side
  applications.
\newblock In {\em Proceedings of the 16th Annual Middleware Conference},
  Middleware '15, pages 75--87, New York, NY, USA, 2015. ACM.

\bibitem{transactionchains}
Y.~Zhang, R.~Power, S.~Zhou, Y.~Sovran, M.~K. Aguilera, and J.~Li.
\newblock {Transaction Chains: Achieving Serializability with Low Latency in
  Geo-distributed Storage Systems}.
\newblock In {\em Proc.\ 24th ACM Symp.\ on Operating Systems Principles}, SOSP
  '13, pages 276--291, New York, NY, USA, 2013. ACM.

\end{thebibliography}

\end{document}